\documentstyle[aps,prd,times]{revtex}

\newcommand{\LieD}{\mbox{\pounds}}

\newcommand{\cE}{{\mathcal{E}}}
\newcommand{\cI}{{\mathcal{I}}}

\begin{document}

\title{A quasilocal calculation of tidal heating}

\author{Ivan S. Booth}
\address{Department of Physics, University of Waterloo, Waterloo, Ontario,
  Canada N2L 3G1}

\author{Jolien D. E. Creighton}
\address{Department of Physics, University of Wisconsin--Milwaukee,
  P.O.~Box 413, Milwaukee, Wisconsin 53201}

\date{1 March 2000}

\wideabs{\maketitle
\begin{abstract}\quad
We present a method for computing the flux of energy through a closed surface
containing a gravitating system.  This method, which is based on the
quasilocal formalism of Brown and York, is illustrated by two applications: a
calculation of (i) the energy flux, via gravitational waves, through a surface
near infinity and (ii) the tidal heating in the local asymptotic frame of a
body interacting with an external tidal field. The second application
represents the first use of the quasilocal formalism to study a non-stationary
spacetime and shows how such methods can be used to study tidal effects in
isolated gravitating systems.
\end{abstract}
\pacs{PACS number(s): 04.20.Cv, 04.25.-g, 04.40.-b, 04.30.-w}}
\narrowtext

\section{Introduction}
\label{s:introduction}

In many physical problems in gravitation, one is interested in the
interaction of a nearly isolated gravitating system with an external
universe.  The interaction effects are computed in a ``buffer zone'' (see
Sec.~20.6 of Ref.~\cite{misner:1973} and Sec.~IB of
Ref.~\cite{thorne:1985}) surrounding the gravitating system, in which the
radius of curvature, scale of inhomogeneity, and rate of change of
curvature are much smaller than the size of the body.  The formalism of
Thorne and Hartle~\cite{thorne:1985} and Zhang~\cite{zhang:1985} has been
used recently by Purdue~\cite{purdue:1999} and Favata~\cite{favata:2000}
to compute the gauge-invariant heating of a body interacting with an
external tidal field.

Until now, calculations of the sort described in these references have
made use of pseudotensors to compute energy and momentum fluxes. However,
quasilocal methods should be equally applicable in situations with a
reasonably well defined buffer zone---in this case, the quasilocal surface
can be conveniently located in the buffer zone. While quasilocal methods
are not fundamentally different than pseudotensor
methods~\cite{chang:1999}, an advantage of quasilocal method is that all
quantities (e.g., energy fluxes) can be computed in terms of real tensors
on the quasilocal surface. Gauge ambiguities in the total amount of energy
and energy flux such as those reported in Ref.~\cite{thorne:1985} and
discussed in Ref.~\cite{purdue:1999} still exist for the quasilocal
methods, but now the ambiguities can be understood in terms of distortions
of the quasilocal surface and so their geometric origin is identified.

In this paper we present a quasilocal formalism for computing the work
done on a gravitating system by an external universe.  Our formalism is
based on the quasilocal mass of Brown and York~\cite{brown:1993}---the
on-shell value of the gravitational Hamiltonian---which coincides with the
Arnowitt-Deser-Misner energy at spatial infinity and the
Trautman-Bondi-Sachs energy at null infinity~\cite{brown:1993,brown:1997}.
It is complementary to but independent of \cite{booth:99} which studied
how motion of the observers affects the Brown-York energy. We use our
expression for energy flux to compute (i) the energy lost in gravitational
radiation from a gravitational system and (ii) the heating of a body
through interactions with an external tidal field.  Problem (i)
demonstrates that the formula for the work reproduces the known
gravitational radiation flux formula when the quasilocal surface is
located in the wave-zone.  Problem (ii) reproduces the calculation of
Purdue~\cite{purdue:1999} using quasilocal methods and shows how these
methods are applicable for problems in which the quasilocal surface is
located in a buffer zone.

\section{Quasilocal energy flux}
\label{s:quasilocalEnergyFlux}

In this section, we derive an expression for the energy flux through a closed
two-surface surrounding a gravitating system.  Our analysis closely follows
Sec.~V of Ref.~\cite{brown:1993}, which derives a conserved measure of mass
for stationary systems.  We relax the requirement that the quasilocal
two-surface time evolution vector be a Killing vector of the spacetime and
thereby obtain an expression for the rate of change in the mass of the system.

Consider a gravitating system separated from the external universe by a
($2+1$)-dimensional timelike boundary $B$.  This boundary has an outward
``radial'' normal vector $n^a$, a metric $\gamma_{ab}=g_{ab}-n_an_b$ induced
by its embedding in the spacetime with metric $g_{ab}$, and an extrinsic
curvature $\Theta_{ab}=-\case{1}{2}\LieD_n\gamma_{ab}$ (with trace
$\Theta=\gamma^{ab}\Theta_{ab}$).  Let $\triangle_a$ be the derivative
operator compatible with the metric $\gamma_{ab}$.  Foliate the boundary $B$
into closed two-surfaces $\Omega_t$ of constant time $t$; then the time
evolution vector $t^a$ on $B$ satisfies $t^a\triangle_at = 1$ and can be
decomposed into a lapse function $N$ and a shift vector $V^a$ on $\Omega_t$ via
$t^a=Nu^a+V^a$, where $u^a$ is the timelike normal to $\Omega_t$
embedded in
$B$.  The closed, spacelike, two-surface $\Omega_t$ has an 
induced metric
$\sigma_{ab}=\gamma_{ab}+u_au_b$ and, viewed as a two-surface embedded in a
three-dimensional spacelike hypersurface $\Sigma$ locally defined such that
$n^a \in T \Sigma$, the extrinsic curvature of $\Omega_t$ is
$k_{ab}=-\case{1}{2}\LieD_n\sigma_{ab}$. 
A full discussion of the geometry 
of the boundary $B$ and its foliation (including a diagram) may be
found in \cite{booth:99}. The notation there is substantially the same
as here though $u^a$ is written as $\tilde{u}^a$.

The Codazzi identity,
\begin{equation}
\label{e:codazzi}
  \triangle_a\tau^{ab}=\gamma^{bc}n^dR_{cd}/8\pi,
\end{equation}
where $\tau^{ab}=(\Theta\gamma^{ab}-\Theta^{ab})/8\pi$, relates the extrinsic
curvature of $B$ to the spacetime Ricci curvature $R_{ab}$.  It then follows
from the Einstein field equations that
\begin{equation}
\label{e:conservation}
  \triangle_a(t_b\tau^{ab}) = t^an^bT_{ab}
  + \case{1}{2}\tau^{ab}\LieD_t\gamma_{ab}.
\end{equation}
We restrict our attention to a vacuum spacetime in which the stress-energy
tensor $T_{ab}$ vanishes.  Then, if $t^a$ is a Killing vector field of the
boundary metric $\gamma_{ab}$, Eq.~(\ref{e:conservation}) is a conservation
equation and the quantity
\begin{equation}
\label{e:mass}
  M = \int_{\Omega_t} d^2x \sqrt{\sigma}\, u_at_b\tau^{ab}
\end{equation}
is a conserved measure of the total mass contained within the boundary
$\Omega_t$.  It is the ``non-orthogonal'' 
Brown-York mass~\cite{brown:1993,booth:99}, 
up to a subtraction term that is required for it to be bounded for large
surfaces in asymptotically flat spacetimes (see, e.g., Lau or Mann
\cite{laurobb}).

When $t^a$ is \emph{not} a Killing vector of the boundary, then 
Eq.~(\ref{e:conservation}) represents an energy flow from the system.
Between two times $t_1$ and $t_2$ one can integrate to find that
$\Delta M = 
-\frac{1}{2}\int_B d^3x \sqrt{-\gamma}\, \tau^{ab}\LieD_t\gamma_{ab}$
is the change in the mass contained by $\Omega_t$. 
Subtraction terms from
a reference spacetime do not need to be included here as it expresses
the \emph{change} in the mass of the system.  The rate at which this
work is done is
\begin{equation}
\label{e:work}
  \frac{dW}{dt} = -\frac{1}{2}\int_{\Omega_t} d^2x \sqrt{-\gamma}\,
  \tau^{ab} \LieD_t\gamma_{ab}
\end{equation}
which describes the rate of change of the system's
mass due to the purely gravitational interaction between it and the surrounding environment.

It is illustrative to decompose the expression for the work into terms
involving projections of $\LieD_t\gamma_{ab}$ normal to and into the spatial
two-surfaces $\Omega_t$.  We find
\begin{equation}
\label{e:workDecomposition}
  \frac{dW}{dt} = \int_{\Omega_t} d^2x \sqrt{\sigma}\,
  \{\case{1}{2}s^{ab}\LieD_t\sigma_{ab}
  - \varepsilon\LieD_tN + j_a\LieD_tV^a\}
\end{equation}
where
$\varepsilon = \sigma^{ab}k_{ab}/8\pi$, 
$j_a = \sigma_{ab}u_c\Theta^{bc}/8\pi$, and  
$s^{ab} = [k^{ab} + \sigma^{ab}
(n^cu^d\triangle_du_c - \sigma^{cd}k_{cd})]/8\pi$
are the quasilocal surface energy, momentum, and stress densities.  
The first two are potentials conjugate to changes in the lapse function
and shift vector respectively while the surface stress density is a work
potential conjugate to changes in the size and shape of the surface
$\Omega_t$.

The stress density can be further decomposed as follows.
A change in the two-metric,
$\delta\sigma_{ab}=\varsigma_{ab}\delta\sqrt\sigma
 +\sqrt\sigma\delta\varsigma_{ab}$, 
is written as a change in the ``size''
$\sqrt\sigma$ of the surface plus a change in the conformally-invariant part of the metric (the ``shape'' of the surface)
$\varsigma_{ab}=\sigma_{ab}/\sqrt{\sigma}$.  Correspondingly, the
surface stress density is decomposed into a surface tension $s=s^{ab}\varsigma_{ab}$
and a shear $\eta^{ab}=s^{ab}/\sqrt{\sigma}$.  Then we rewrite the 
work term as
$\case{1}{2}s^{ab}\LieD_t\sigma_{ab}=\case{1}{2}(s\LieD_t\sqrt\sigma
+\eta^{ab}\LieD_t\varsigma_{ab})$.

The above has a particularly nice application in the physics of thin shells.
Israel~\cite{israel:1966} first showed that a thin shell of matter can be
described in general relativity by matching two spacetimes along a timelike
boundary $B$ such that even though they induce the same surface metric on $B$,
the extrinsic curvature in each spacetime is different. If $\Theta^+_{ab}$ and
$\Theta^-_{ab}$ are those curvatures this (mild) singularity can be accounted
for if there is a (distributional) stress energy tensor
$S_{ab}=\tau^+_{ab}-\tau^-_{ab}$ over $B$.  A set of observers dwelling on the
surfaces $\Omega_t$ (which foliate $B$) measures the shell to have
matter-energy $M^+-M^-$ [Eq.~(\ref{e:mass})]. A more detailed discussion of
this may be found in \cite{booth:99} but here we note that the above analysis
for the quasilocal energy also shows that the set of observers dwelling on
$\Omega_t$ measures the matter-energy to change with rate $dW/dt$
(Eq.~\ref{e:workDecomposition}).  Then the quasilocal densities defined above
are the energy, angular momentum, and stress tensor of the \emph{matter
shell}. A set of observers being evolved by $t^a = u^a$ see work being done on
the shell at a rate equal to the integral of the stress tensor contracted with
the time rate of change of the area---exactly as one would expect from
classical physics.

\section{Gravitational radiation}
\label{s:gravitationalRadiation}

Equation~(\ref{e:work}) purportedly measures the change in the mass of a
system.  In this section we apply our work formula to obtain the correct mass loss for a system radiating gravitational waves.  For this
we suppose that the quasilocal surface is in the wave-zone, far away
from the radiating system.  Although this is not a very interesting
application of a quasilocal method (since an asymptotic method, such 
as the Bondi-Sachs mass loss formula could as well be used), it is
useful to confirm that Eq.~(\ref{e:work}) does recover the correct
result.

Gravitational radiation far from the generating source can be 
described as a transverse-traceless perturbation to the flat-space
metric. In spherical-polar coordinates, the metric is given by
$ds^2=-dt^2+dr^2+(rd\theta)^2+(r\sin\theta d\phi)^2+h_{\mu\nu}dx^\mu dx^\nu$ where
$h_{\mu\nu}dx^\mu dx^\nu=h_+[(rd\theta)^2-(r\sin\theta d\phi)^2]
 +2h_\times(rd\theta)(r\sin\theta d\phi)$
is the transverse, trace free perturbation.  The ``plus,'' $h_+$, and
``cross,'' $h_\times$, polarizations represent outgoing, spherical waves, and
have the form $h_+(t,r,\theta,\phi)=s_+(t-r,\theta,\phi)/r$ and
$h_\times(t,r,\theta,\phi)=s_\times(t-r,\theta,\phi)/r$.  We then find the energy lost by the radiating system by inserting this metric into
Eq.~(\ref{e:work}) while taking the boundary to be a sphere of constant
$r$ in the wave-zone (very large $r$).  The integrand of
Eq.~(\ref{e:work}) is
\begin{equation}
\label{e:waveFlux}
  \frac{dE}{dt\,d\Omega_t} = -\frac{r^2}{16\pi}[(\partial h_+/\partial t)^2
  + (\partial h_\times/\partial t)^2]
\end{equation}
to leading order in the perturbation and in $r$.  This is the standard
expression for the flux of gravitational radiation---see, e.g., Eq.~(10) of
Ref.~\cite{thorne:1987}.

By inspection of the form of the perturbation, it is clear that the energy
loss arises due to the shearing of the bounding two-surface $\Omega_t$ since,
to leading order, the perturbation does not affect the volume element on that
two-surface.  Thus the entire energy loss (in the transverse, trace-free
gauge) arises from the ``$\eta^{ab}\LieD_t\varsigma_{ab}$'' work term.

As a simple example, consider two point-particles, each of mass $m=M/2$,
orbiting each other in the $xy$-plane with angular frequency $\omega$ and
constant separation $a$.  The quadrupole moment tensor $\cI_{jk}$
in Cartesian coordinates is
$\cI_{xx}=-\cI_{yy}=\case{1}{8}Ma^3\cos2\omega t$ and
$\cI_{xy}=\case{1}{8}Ma^2\sin2\omega t$ (constant terms omitted).  The
far-field metric perturbation is $h_{jk}=2(\partial^2\cI_{jk}/\partial
t^2)/r$, so $h_{yy}=-h_{xx}=(Ma^2\omega^2/r)\cos2\omega(t-r)$ and
$h_{xy}=-(Ma^2\omega^2/r)\sin2\omega(t-r)$.

Using Eqs.~(4.3) and~(4.4) of Ref.~\cite{wiseman:1992}, we find
\begin{mathletters}
\begin{eqnarray}
  h_+ &=& -\case{1}{2}Ma^2\omega^2r^{-1}(1+\cos^2\theta)\cos2[\omega(t-r)-\phi] 
  \\
  h_\times &=& -Ma^2\omega^2r^{-1}\cos\theta\sin2[\omega(t-r)-\phi].
\end{eqnarray}
\end{mathletters}%
We then integrate Eq.~(\ref{e:waveFlux}) over the sphere at large $r$ to
obtain the loss of energy from the system:
\begin{equation}
  -dE/dt = \case{2}{5}M^2a^4\omega^6 = \case{2}{5}(M/a)^5
\end{equation}
where we have used Kepler's law $a^3\omega^2=M$ for particles in a circular
orbit.

\section{Tidal heating}
\label{s:tidalHeating}

We now calculate the work done by an external gravitational field to deform a
self-gravitating body.  The canonical example of this effect in the solar
system is the tidal heating of Io by Jupiter.  In this instance, the gradient
of Jupiter's gravitational field distorts Io from being a perfect sphere and
then tidally locks it in its orbit so that it always presents the same face to
Jupiter.  That orbit is strongly perturbed by the other Gallilean moons and so
its radial distance from Jupiter varies with time.  With this variation comes
a corresponding one in the gradient of the field and so Io is gradually
stretched and then allowed to relax.  The energy transferred by this pumping
is largely dispersed as heat and it is this heat that produces the volcanic
activity on Io.  The same type of process occurs for any two bodies in
non-circular orbits about each other.

First from a Newtonian perspective, we may mathematically describe the
gravitational fields in this situation as follows.  We assume that the
self-gravitating body is far enough away from the source of the external field
that that field is nearly uniform close to the body.  Then in a rectangular
coordinate system that orbits with the body with its origin at the center of
mass, the Newtonian potential of the external field may be written as
$\Phi_{\text{ext}}=\case{1}{2}\cE_{ij}x^ix^j$ where $\cE_{ij}$ 
is the (time-dependent but
symmetric and trace-free) quadrupole moment of the field and $x^i$ is the
position vector based at the body's centre of mass. At the same time, to
quadrupolar order the Newtonian potential of the body is
$\Phi_{\text{o}}=-M/r-\case{3}{2}r^{-3}\cI_{ij}n^in^j$, where $M$ is the mass
of the body, $r$ is the radial distance from the centre of mass, $\cI_{ij}$ is
its (time-dependent but symmetric and trace-free) quadrupole moment, and $n^i$
is the unit normal radial vector. 

With this in mind the techniques of Thorne and Hartle \cite{thorne:1985} can
be used to construct a metric that describes these situations in the slow
moving, nearly Newtonian limit.  First, define an annulus surrounding the body
whose inner boundary is chosen so that the gravitational field of the body is
weak throughout and whose outer boundary is chosen so that the external field
is nearly uniform.  This region is termed the buffer zone.  The rectangular
coordinate system is replaced with one that is chosen so that the metric is as
close to Minkowskian as possible over the buffer zone \cite{purdue:1999}.
Then to first order in perturbations from Minkowski and first order in time
derivatives the metric can be written as
\begin{eqnarray}
  ds^2 &=& - (1+2 \Phi) dt^2 + 2(A_j + \partial_t\xi_j) dx^j dt \nonumber\\ 
  && + [(1-2\Phi) \delta_{ij} + \partial_i\xi_j + \partial_j\xi_i] dx^i dx^j
\end{eqnarray}
where the indices run from one to three and $\delta_{ij}$ is the Cartesian
metric ${\mathrm{diag}}[1,1,1]$ on a spacelike slice.  The Newtonian potential
is $\Phi=-M/r-\case{1}{2}(3r^{-3}\cI_{ij}-r^2\cE_{ij})n^i n^j$ and
$A_j=-2r^{-2}n^kd\cI_{jk}/dt-\case{2}{21}r^3(5n_jn^k-2\delta^k_j)n^l
 d\cE_{kl}/dt$ is a vector potential that must be added so that the metric is
a solution to the first order Einstein equations.  Here, $n^i$ is the radial
normal with respect to the flat spatial metric $\delta_{ij}$ and
$r^2=x^2+y^2+z^2$.  The diffeomorphism generating vector field $\xi_j$
represents the gauge ambiguity in setting up a nearly Minkowski coordinate
system.  In order that the metric be slowly evolving and nearly Minkowski,
$\xi_j$ must be of the form
$\xi_j=\alpha r^{-2}\cI_{jk}n^k+\beta r^3 \cE_{jk}n^k+\gamma r^3\cE_{kl}n^kn^ln_j$,
where  $\alpha$, $\beta$, and $\gamma$ are free constants of order one. 

We set up a constant $r$ timelike quasilocal surface $B$ in the buffer zone
and foliate with constant $t$ spacelike two-surfaces $\Omega_t$.  Then the
time vector $t^a$ is $\partial/\partial t$. In calculating the rate of change
of the mass contained within $\Omega_t$ it is most convenient to switch to
spherical coordinates.  We make the standard transformation to spherical
coordinates $x^i=r[\sin\theta\cos\phi,\sin\theta\sin\phi,\cos\theta]$; in
these coordinates, the metric is
\begin{eqnarray}
ds^2 &=& -(1+2\Phi)dt^2 + (1-2\Phi)[dr^2 + (rd\theta)^2 \nonumber\\
  && + (r\sin\theta d\phi)^2] + 2\bar{A}_r dr dt + 2\bar{A}_\theta (rd\theta)dt
  \nonumber\\
  && + 2\bar{A}_\phi (r\sin\theta d\phi)dt + H_{rr}dr^2 
  + H_{\theta\theta}(rd\theta)^2 \nonumber\\
  && + H_{\phi\phi}(r\sin\theta d\phi)^2 + 2H_{r\theta}dr(rd\theta) \nonumber\\
  && + 2H_{r\phi}dr(r\sin\theta d\phi)
  + 2H_{\theta\phi}(rd\theta)(r\sin\theta d\phi)^2
\end{eqnarray}
where 
$H_{rr}=-4\alpha r^{-3}\cI_{rr}+6(\beta+\gamma)r^2\cE_{rr}$,
$H_{\theta\theta}=2\alpha r^{-3}\cI_{\theta\theta}+2\beta r^2\cE_{\theta\theta}
 +2\gamma r^2\cE_{rr}$,
$H_{\phi\phi}=2\alpha r^{-3}\cI_{\phi\phi}+2\beta r^2\cE_{\phi\phi}
 +2\gamma r^2\cE_{rr}$,
$H_{r\theta}=-\alpha r^{-3}\cI_{r\theta}+(4\beta+2\gamma)r^2\cE_{r\theta}$,
$H_{r\phi}=-\alpha r^{-3}\cI_{r\phi}+(4\beta+2\gamma)r^2\cE_{r\phi}$, and
$H_{\theta\phi}=2\alpha r^{-3}\cI_{\theta\phi}+2\beta r^2\cE_{\theta\phi}$.  
In these expressions $\cE_{rr}=\cE_{ij}e^i_re^j_r$,
$\cE_{r\theta}=\cE_{ij}e^i_re^j_\theta$, etc., with $e^i_r=n^i$,
$e_\theta^i=\partial_\theta e^i_r$ and
$e_\phi^i=(1/\sin\theta)\partial_\phi e^i_r$.  Also,
$\bar{A}_r=(A_j+\partial_t\xi_j)e^j_r$,
etc., but we don't need their expanded forms since only time derivatives of
them show up in later calculations and we are ignoring second order time
derivatives.

As might be expected, the subsequent calculations are quite involved and we did
them partially with GRTensor~\cite{grtensor}.  To lowest order
\begin{eqnarray}
\label{e:power1}
  \frac{dW}{dt} &=& -\frac{1}{2}\int_{\Omega_t} d^2x \sqrt{-\gamma}\,
  \tau^{ab} \LieD_t\gamma_{ab} \nonumber \\
  &=& \frac{1}{2} \cE_{ij} \frac{d\cI_{ij}}{dt}
  + \frac{1}{60} \frac{d}{dt} [ 2(-3 - 2\beta - 2\beta^2 + 4\gamma \nonumber\\
  &&\quad + 4\gamma^2 + 8 \beta \gamma ) r^5 \cE_{ij} \cE_{ij} \nonumber\\
  &&\quad + 2(3 - 2\alpha + 6\beta - 12\gamma + 8\alpha\gamma) \cE_{ij} \cI_{ij}
  \nonumber \\
  &&\quad - (-9 + 12\alpha + 4\alpha^2) r^{-5} \cI_{ij} \cI_{ij} ].
\end{eqnarray}
The calculations used the identities 
$\int_{\Omega_t} d\theta d\phi\sin\theta A_{rr}B_{rr}=(8\pi/15)A_{ij}B_{ij}$ and
$\int_{\Omega_t} d\theta d\phi\sin\theta(2A_{\theta\phi}B_{\theta\phi}
 -A_{\theta\theta}B_{\phi\phi}-A_{\phi\phi}B_{\theta\theta})
 =(4\pi/3)A_{ij}B_{ij}$ where the integrations are over the unit sphere.

This result requires some interpretation.  As the external field changes with time and thereby forces the self-gravitating body to change configuration, the
work done by the external field can be split into time reversible and
irreversible parts [as seen in Eq.~(\ref{e:power1})].  The reversible work
represents work being done to increase the potential energy of the system and
is recoverable.  On the other hand the irreversible part represents work done
to deform and/or heat up the system.  This is the tidal heating that we are
interested in.  Further, from the quasilocal perspective, we expect to see an
energy flow arising from fluctuations of the quasilocal surface within
otherwise static fields.  Of course this work would also be reversible.  Thus,
it is only the irreversible part that we are interested in and we have
calculated that to be $\case{1}{2}\cE_{ij}d\cI_{ij}/dt$ above.  This is the same leading term obtained when one does the corresponding calculation in Newtonian gravity or with pseudotensors
\cite{purdue:1999} and it is independent of diffeomorphisms 
generated by $\xi_j$ which correspond to fluctuations of the quasilocal
surface. Note however, that the time reversible and 
gauge dependent terms of equation (\ref{e:power1}) are dependent on
those fluctuations and furthermore that dependence is different from
that found in ref.~\cite{purdue:1999} using pseudo-tensor methods.
Similarly other pseudo-tensor or quasilocal methods
would obtain a different gauge dependence for these terms. What is
important is that the physically relevant time irreversible term does
not depend on the $\xi_j$-generated diffeomorphisms. 

Finally for completeness let us consider how this energy flow splits up into
its components parts as considered in Eq.~(\ref{e:workDecomposition}).  Then
to the order that we are interested the angular momentum term is zero and we
are left with two terms
$d{W}_N/dt=-\int d\theta d\phi\sqrt{\sigma}\varepsilon\LieD_tN$ and
$d{W}_\sigma/dt=\case{1}{2}\int d\theta d\phi
 \sqrt{\sigma}Ns^{ab}\LieD_t\sigma_{ab}$.  We find
\begin{eqnarray}
  \frac{dW_N}{dt} &=&
  \frac{1}{2} \cE_{ij} \frac{d\cI_{ij}}{dt}
  + \frac{\alpha}{15} \frac{d\cE_{ij}}{dt} \cI_{ij}
  - \frac{\beta}{5} \cE_{ij} \frac{d\cI_{ij}}{dt}
  - \frac{4\gamma}{5} \cE_{ij} \frac{d\cI_{ij}}{dt} \nonumber \\
  && + \frac{1}{60} \frac{d}{dt} [
  2(4\gamma + \beta - 2) r^5 \cE_{ij}\cE_{ij}
  - 6\cE_{ij} \cI_{ij} \nonumber\\
  &&\qquad - 3(2 \alpha - 3)r^{-5} \cI_{ij}\cI_{ij}].
\end{eqnarray}
The second term is a bit more complicated. It is
\begin{eqnarray}
  \frac{dW_\sigma}{dt} &=& 
  - \frac{\alpha}{15} \frac{d\cE_{ij}}{dt} \cI_{ij}
  + \frac{\beta}{5} \cE_{ij} \frac{d\cI_{ij}}{dt}
  + \frac{4\gamma}{5} \cE_{ij} \frac{d\cI_{ij}}{dt} \nonumber\\
  && + \frac{1}{30}\frac{d}{dt} [ (-1 - 3\beta - 2\beta^2 + 4\gamma^2
  + 8 \beta \gamma) r^5\cE_{ij} \cE_{ij} \nonumber \\
  &&\qquad + 2(3 - \alpha + 3\beta - 6\gamma 
  + 4\alpha\gamma) \cE_{ij} \cI_{ij} \nonumber  \\
  &&\qquad - (2\alpha^2 - 9\alpha + 9)
  r^{-5} \cI_{ij} \cI_{ij} ].
\end{eqnarray}
Thus part of the work done is measured by deformations of the surface and part
is measured by changes in how observers choose to measure the rate of passage
of time.  Note that individually the time irreversible sections of the two
parts are gauge dependent, but when we combine them we reobtain
Eq.~(\ref{e:power1}) and the gauge dependence vanishes back into the
reversible part where we would expect it.

\section{Conclusions}
\label{s:conclusions}
We have modified the quasilocal energy formalism of Brown and York so
that it may be used to study non-stationary spacetimes where energy 
flows in and out through the quasilocal surface. 
As applications of this extension we have examined implications for
the physics of relativistic thin shells of matter, the energy carried
from a source to infinity by gravitational waves, and the transfer of
energy to a body during gravitational tidal heating. The success
of the formalism in all three applications provides further evidence
that the Brown-York energy has physical content. Furthermore, in 
the tidal heating application we have seen how the quasilocal formalism
provides a geometrical explanation of the gauge ambiguities that are
also found in the Newtonian and pseudotensor approaches.

\acknowledgements
The authors would like to thank Robert Mann, Patricia Purdue, Alan Wiseman, and Kip Thorne
for their useful comments and suggestions.  This work was supported by the
Natural Sciences and Engineering Research Council of Canada and NSF grants PHY-9728704 and AST-9731698.

\end{document}